\RequirePackage{fix-cm}
\documentclass{svjour3}                     
\smartqed  
\usepackage{graphicx}
\usepackage[colorlinks,citecolor=red]{hyperref}
\usepackage{amsmath}
\begin{document}

\title{Hyperspherical approach to atom--dimer collisions with the Jacobi boundary condition
}


\author{Cai-Yun Zhao \and  Yi Zhang \and  Hui-Li Han \and  Ting-Yun Shi
}


\institute{Cai-Yun Zhao \and  Yi Zhang \and  Hui-Li Han \and  Ting-Yun Shi \at
              State Key Laboratory of Magnetic Resonance and Atomic and Molecular Physics, Wuhan Institute of Physics and Mathematics, Innovation Academy for Precision Measurement Science and Technology, Chinese Academy of Sciences, Wuhan 430071, P. R. China \\
           \and
           Cai-Yun Zhao \and  Yi Zhang \at
              University of Chinese Academy of Sciences, 100049, Beijing, P. R. China \\
           \and
           Hui-Li Han \at
              \email{huilihan@wipm.ac.cn}
}

\date{Received: date / Accepted: date}

\maketitle

\begin{abstract}
In this study, we investigate atom--dimer scattering within the framework of hyperspherical coordinates. The coupled-channel Schr\"odinger equation is solved using the R-matrix propagation technique combined with the smooth variable discretization method. In the matching procedure, the asymptotic wave functions are expressed in the rotated Jacobi coordinates. We apply this approach to the elastic scattering $^{3}$He(T$\uparrow$) + $^{4}$He$_{2}$ and H$\uparrow$ +LiH$\uparrow$ processes for testing. The convergence of the scattering length as a function of the propagation distance is studied. We find that the method is reliable and can provide considerable savings over previous propagators, so it is suitable for solving the atom--dimer scattering problem for important quantities such as the phase shift, cross section and scattering length.
\end{abstract}

\section{Introduction}
\label{intro}
Studies of three-body collision processes have attracted tremendous attention due to their substantial relevance in the rapidly growing field of cold and ultracold atomic gases \;\cite{PhysRevLett.90.053202,PhysRevLett.103.083202,RevModPhys.89.035006,Giannakeas2017Van}. In such systems, elastic atom--molecule collisions are crucial for determining the dynamics of ultracold atom--molecule mixtures at the mean field level, and inelastic atom--molecule collisions have a large impact on the lifetime of Feshbach molecules.

Weakly attracted three-body systems such as helium trimer and mixed $^{4}$He-$^{4}$He-A (A is another atom) systems are very interesting and important as they give us an opportunity to study the Efimov states in the realistic systems\;\cite{2014Imaging,PhysRevA.79.024501,Wumengshan2014}. Interest in the Efimov states and other universal binding properties of such systems has been significantly investigated, and giant Efimov trimer has been detected in helium gas\;\cite{Kunitski2015He}. The scattering processes at ultralow energies are even more interesting due to their relevance for the lifetime and stability of gas samples. A few works have addressed the ultracold atom--molecule problem in these realistic systems. For instance, ultracold collisions of $^{3,4}$He atoms with $^4$He$_{2}$ have been studied within the adiabatic hyperspherical representation by Refs.\;\cite{PhysRevA.78.062701,PhysRevA.83.032703}. The Faddeev differential equations have also been extensively used in these systems\;\cite{2004Binding,2006Scattering,Kolganova2009Ultracold}. In addition to the well-studied elastic $^4$He($^{3}$He)+$^4$He$_{2}$ scattering, the spin-stretched case of H atom scattering from XH (X is an alkali atom) has been investigated using the method of hyperspherical coordinates\;\cite{PhysRevA.83.032703}. Recently, atom--dimer exchanges and dissociation reaction rates have been predicted for different combinations of two $^{4}$He atoms and one of the alkaline species among $^6$Li, $^7$Li, and $^{23}$Na using the Faddeev formalism\;\cite{PhysRevA.102.062814}. On the other hand, it is known that there exist many similarities between spin-polarized tritium (T$\uparrow$) and $^{4}$He atoms. The bulk T$\uparrow$ remains liquid in the limit of zero temperature and behaves much like liquid $^{4}$He and therefore constitutes a second example of a bosonic superfluid. The bound states of mixed T$\uparrow^{4}$He$_{2}$ clusters were studied in Refs.\;\cite{PhysRevLett.113.253401,Hiroya2014A,doi:10.1063/1.3530837} and were found to possess one weakly bound state, which is by far the most weakly bound system. For this system, no scattering observables are available in the literature, which is of fundamental importance for current experiments.

The hyperspherical adiabatic (HA) expansion method has been proven to be an efficient tool in studying few-atom systems\;\cite{CDLIN1995}. For bound states, HA expansion shows particularly fast convergence for atom--atom interactions\;\cite{Wumengshan2014,doi:10.1063/1.3451073,PhysRevA.79.024501,PhysRevA.84.014501}. On the other hand, the method has also been extensively used to describe few-atom systems in the ultracold collision regime\;\cite{PhysRevLett.90.053202,PhysRevA.78.062701,PhysRevA.83.032703}. The convergence problem of the HA method appears for scattering states, particularly in the description of ultracold atom--dimer collisions. Since the asymptotic structure for atom--dimer scattering is that one particle moves relative to the center of mass of the two-body bound system, the correct boundary condition for the structure with the HA basis is achieved only at $\rho\rightarrow\infty$, which requires a very large number of hyperradial functions in the solutions and long-range
propagation\;\cite{PhysRevLett.103.090402}. To overcome the convergence problem, Refs.\;\cite{PhysRevLett.103.090402,PhysRevA.83.022705,2008Three} introduced a method to compute the phase shift from two integral relations that involve only the internal part of the wave function. The convergence of the procedure has been demonstrated to be as fast as for bound states.

An alternative method for addressing this problem is using asymptotic solutions expressed in Jacobi coordinates. This idea has been applied to treat rearrangement collisions by several quantum chemistry groups since 1980\;\cite{doi:10.1063/1.476337}. In the calculations of Refs.\;\cite{DUEBILLING1980254,BONDI1982570,doi:10.1021/j150644a017}, the probabilities were found to exhibit no oscillations as a function of the matching distance. Then, in treating the collision-induced dissociation problem, Refs.\;\cite{doi:10.1063/1.1573186,Parker} used a mixed boundary condition scheme, in which the asymptotic bound solutions were expressed in Jacobi coordinates and the continuum solutions were expressed in hyperspherical coordinates. This method significantly decreases the amplitude of the oscillations and improves the convergence as a function of the distance. In atomic physics, the hyperspherical close-coupling method, which uses Jacobi asymptotic solutions, has been used to calculate the elastic and positronium formation cross section for electron and positron collisions with atomic hydrogen\;\cite{1994,Zhou_1995,PhysRevA.50.232} and to study the photoionization cross section spectra of the two-electron system\;\cite{Zhou_19931,Zhou_19932}. Zhao \textit{et al}.\;\cite{PhysRevA.62.042706} also used the hyperspherical close-coupling method to investigate the charge transfer process $A^{+} + B \rightarrow A + B^{+} $. In the calculations of the low-energy collision of Coulomb three-body systems, Refs.\;\cite{PhysRevA.92.032713,doi:10.1063/1.476337} used the hyperspherical elliptic coordinates method. Their two-dimensional matching procedure also used the asymptotic wave function expressed in the mass-scaled Jacobi coordinates. For the ultracold atom--dimer elastic scattering, the collision quantities reach the threshold regime only at collision energies at the $nk$ level, leading to longer propagation than in the systems described above. Thus, there is a great need to project numerical wave functions onto asymptotic solutions expressed in Jacobi coordinates to study the ultracold atom--dimer scattering process.

In this work, we present an efficient method for investigating atom--dimer scattering within the framework of hyperspherical coordinates. The nonadiabatic coupling between the hyperradius and hyperangular variables is treated with the slow-variable discretization (SVD) method\;\cite{Tolstikhin_1996} in combination with the R-matrix propagation technique\;\cite{doi:10.1063/1.432836,PhysRevA.84.052721}. In the matching procedure, the asymptotic wave functions are expressed in the rotated Jacobi coordinates. We perform test calculations on the $^{3}$He + $^{4}$He$_{2}$ and H$\uparrow$ + H$\uparrow$Li systems, which represent two different kinds of asymptotic structures. The two systems have been studied previously with the asymptotic wave function expressed in the hyperspherical coordinates, and the propagation distance is at least 5000 a.u.\;\cite{PhysRevA.83.032703}. Thus, these systems are good examples to illustrate our new approach. We also investigate the T$\uparrow$ + $^{4}$He$_{2}$ elastic scattering in $J^{\Pi}=0^{+}$ symmetries and provide the scattering length value for T$\uparrow$ atom scattering from the $^{4}$He$_{2}$ dimer.

The organization of this paper is as follows: Sec. II describes the theoretical approach. In Sec. III, we discuss the results and analyses of the systems under study. Finally, we conclude and summarize our work in Sec. IV.

\section{Theoretical formalism}
In this work, we consider a process where a particle hits a bound two-body system. We assume the incident energy to be below the breakup threshold for the three particles, and only the channels approaching the two-body bound state need to be considered.

\begin{figure}
\centering
\includegraphics[width=0.6\textwidth]{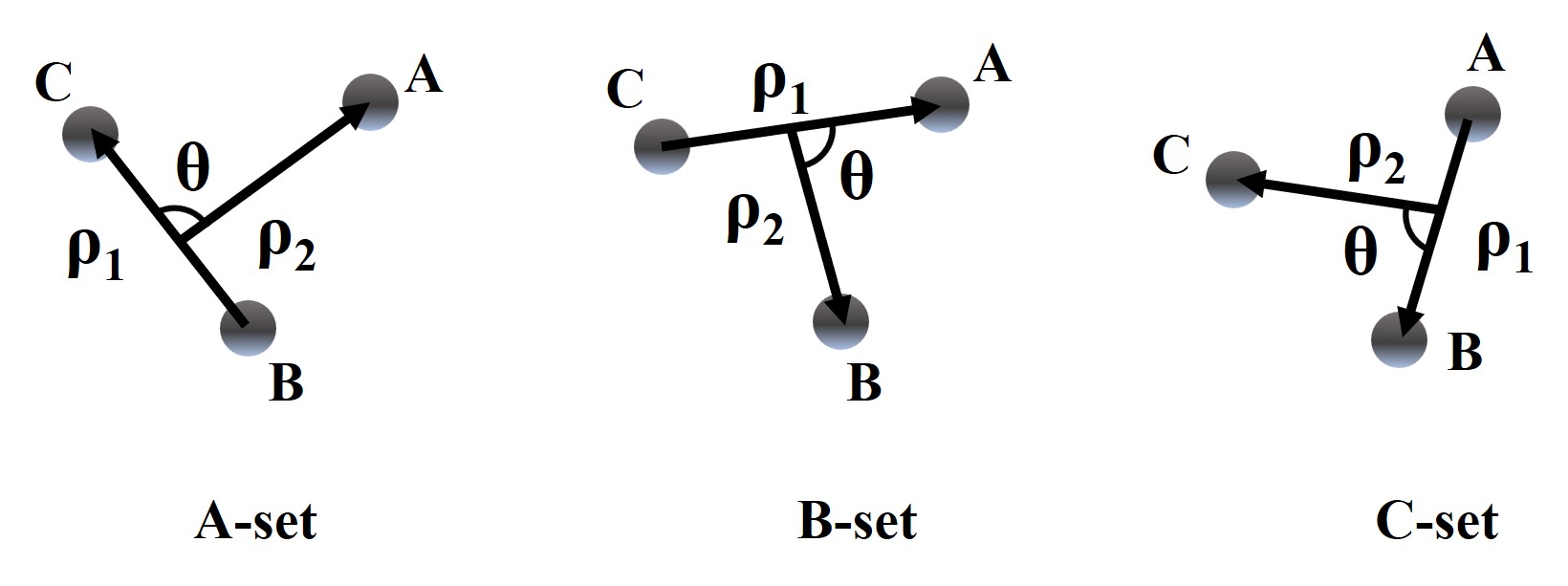}
\caption{Three sets of Jacobi coordinate vectors.}
\label{f2-1}
\end{figure}

We use $m_\tau$ ($\tau$=A, B, C) to represent the mass of three atoms and use $\mathbf{x_\tau}$ to represent the column vector relative to the origin. In the center-of-mass frame, six coordinates are needed to describe the three-particle system.
The Jacobi coordinates describing the relative motion are defined as
\begin{equation}
\begin{array}{l}
\mathbf{\rho}_{2\tau}=\mathbf{x}_{\tau}-\frac{m_{\tau+1} \mathbf{x}_{\tau+1}+m_{\tau+2} \mathbf{x}_{\tau+2}}{m_{\tau+1}+m_{\tau+2}}, \\
\mathbf{\rho}_{1\tau}=\mathbf{x}_{\tau+2}-\mathbf{x}_{\tau+1},
\end{array}
\end{equation}
where $\tau,\tau+1,\tau+2$ are any cyclic permutations of A, B, and C. This is illustrated in Fig. \ref{f2-1}.
In addition, $\mathbf{\xi_{1\tau}}$ and $\mathbf{\xi_{2\tau}}$ are the corresponding mass-scaled Jacobi coordinates:
\begin{equation}
\mathbf{\xi}_{2\tau}=\sqrt{\frac{\mu_2}{\mu}} \mathbf{\rho}_{2\tau}, \quad
\mathbf{\xi}_{1\tau}=\sqrt{\frac{\mu_1}{\mu}} \mathbf{\rho}_{1\tau},
\end{equation}
where $\mu_1=\frac{m_{\tau+1}m_{\tau+2}}{m_{\tau+1}+m_{\tau+2}}$,
$\mu_2=\frac{m_\tau(m_{\tau+1}m_{\tau+2})}{m_\tau+m_{\tau+1}+m_{\tau+2}}$ and $\mu=\left[\frac{m_\tau m_{\tau+1}m_{\tau+2}}{m_\tau + m_{\tau+1}+m_{\tau+2}}\right]^{1 / 2}$.
Different sets of mass-scaled Jacobi coordinates can be transformed through kinematic rotations:
\begin{equation}
\label{tfq}
\begin{aligned}
\left(\begin{array}{l}
\mathbf{\xi_2}_{\tau+1} \\
\mathbf{\xi_1}_{\tau+1}
\end{array}\right)
=\mathbf{T}
\left(\begin{array}{l}
\xi_{2\tau }\\
\xi_{1\tau }
\end{array}\right),
\end{aligned}
\end{equation}
where
\begin{equation}
\label{tfqt}
\mathbf{T}\left(\chi_{\tau+1, \tau}\right)=\left(\begin{array}{cc}
\cos \left(\chi_{\tau+1, \tau}\right) \mathbf{1} & \sin \left(\chi_{\tau+1, \tau}\right) \mathbf{1} \\
-\sin \left(\chi_{\tau+1, \tau}\right) \mathbf{1} & \cos \left(\chi_{\tau+1, \tau}\right) \mathbf{1}
\end{array}\right)
\end{equation}
is a $6\times 6$ matrix and $\mathbf{1}$ is the $3\times 3$ unit matrix.
The kinematic angles $\chi_{\tau+1,\tau}$ are negative and obtuse, revealing the mass of three particles:
\begin{equation}
\begin{array}{l}
\cos \chi_{\tau+1, \tau}=-\frac{\mu}{d_{\tau} d_{\tau+1} m_{\tau+2}}, \\
\sin \chi_{\tau+1, \tau}=-\frac{1}{d_{\tau} d_{\tau+1}},
\end{array}
\end{equation}
where $d_{\tau}=\left[\frac{m_{\tau}}{\mu}\left(1-\frac{m_{\tau}}{m_\tau + m_{\tau+1}+m_{\tau+2}}\right)\right]^{1 / 2}$ is a scaling factor.

Delves hyperspherical coordinates can be defined in any set of mass-scaled Jacobi coordinates. In this work, hyperspherical coordinates are defined in the $A$-set ($\tau=A$) mass-scaled Jacobi coordinates, where two identical atoms are connected through the Jacobi vector $\vec{\rho_1}$. We denote the angle
between $\vec{\rho}_{1}$ and $\vec{\rho}_{2}$ as $\theta$. The channel functions are symmetric with respect to the $\theta$ direction in this definition. After separation of the center of the mass motion,
three of the six coordinates are taken to be the Euler angles---$\alpha$, $\beta$, and $\gamma$---that specify the orientation of the body-fixed frame relative to the space-fixed frame.
The remaining degrees of freedom can be represented by the hyperradius $R$ and the two hyperangles $\theta$ and $\phi$, which are defined as\;\cite{CDLIN1995}
\begin{equation}
\label{1}
\mu R=\mu_1 \rho^2_1+\mu_2 \rho^2_2
\end{equation}
and
\begin{equation}
\label{2}
\tan \phi =\sqrt{\frac{\mu_2}{\mu_1}}\frac{\rho_2}{\rho_1},\;\; 0 \leq \phi\leq\frac{\pi}{2},
\end{equation}
respectively.
$R$ is the only coordinate with the dimension of length, which represents the size of the three-body system.
Here, $\theta$, $\phi$ and the three Euler angles $(\alpha,\beta,\gamma)$ can be collectively represented by $\Omega$ $[\Omega\equiv(\theta,\phi,\alpha,\beta,\gamma)]$. In our method, wave functions are expanded in the body frame $xyz$, where $\vec{\rho_2}$ lies along the $\vec{z}$-axis and the three particles lie on the $xz$ plane.

We introduce the reduced wave function $\psi_{\upsilon'}(R;\Omega)=\Psi_{\upsilon'}(R;\Omega)R^{5/2}\sin\phi\cos\phi$, and the Schr\"odinger equation is of the form:
\begin{equation}
\label{3}
\bigg[-\frac{1}{2\mu}\frac{d^2}{dR^2}+\bigg(\frac{\Lambda^2-\frac{1}{4}}{2\mu R^2}
+V(R;\theta,\phi)\bigg)\bigg]\psi_{\upsilon'}(R;\Omega) =E\psi_{\upsilon'}(R;\Omega)\,,
\end{equation}
where $\Lambda^2$ is the squared ``grand angular momentum operator'', whose expression is given in Ref.\;\cite{CDLIN1995}.
The three-body interaction $V(R;\theta,\phi)$ in Eq.(\ref{3}) is taken to be a sum of the three pairwise two-body interactions.

Equation\;(\ref{3}) is solved in the hyperspherical adiabatic representation. Similar to the usual adiabatic approximation, the hyperspherical adiabatic potentials $U_\nu(R)$ and channel functions $\Phi_\nu(R;\Omega)$ are defined as solutions of the following adiabatic eigenvalue problem:
\begin{equation}
\label{9}
\bigg(\frac{\Lambda^{2}-\frac{1}{4}}{2\mu R^{2}}+
V(R;\theta,\phi)  \bigg)\Phi_{\nu}(R;\Omega)=U_{\nu}(R)\Phi_{\nu}(R;\Omega)\,.
\end{equation}

We define the normalized and symmetrized D-functions associated with our choice of the body frame:
\begin{align}
\tilde{D}^{J\Pi}_{IM}(\alpha,\beta,\gamma)=\frac{1}{4\pi}\frac{\sqrt{2J+1}}{\left[1+(\sqrt{2}-1)\delta_{I0}\right]}
\left[D^J_{IM}+(-1)^{I+J}\Pi D^J_{-IM}\right],
\end{align}
where $J$ is the total nuclear orbital angular momentum, $M$ is its projection onto the laboratory-fixed axis, and $\Pi$ is the parity with respect to the inversion of the nuclear coordinates. The quantum number $I$ denotes the projection of $J$ onto the body-frame $z$ axis.

The channel functions are expanded in terms of D-functions as follows:
\begin{align}
\Phi_\nu^{J\Pi M}(R;\Omega)=\sum^J_{I=0}u_{\nu I}(R;\theta,\phi)\tilde{D}^{J\Pi}_{IM}(\alpha,\beta,\gamma),
\end{align}
and $u_{\nu I}(R;\theta,\phi)$ is expanded with B-spline functions,
\begin{equation}
u_{\nu I}(R;\theta,\phi)=\sum^{N_{\phi}}_i\sum^{N_{\theta}/2}_j
c_{i,j}B_i(\phi)\left[B_j(\theta)+B_{N_{\theta}+1-j}\left(\pi-\theta\right)\right],
\end{equation}
where $N_{\theta}$ and $N_{\phi}$ are the sizes of the basis sets in the $\theta$
direction and $\phi$ direction, respectively. The constructed symmetric B-spline basis sets utilized in the $\theta$ direction reduce the
number of basis functions to $N_{\theta}/2$.

Following the method of Ref.\;\cite{PhysRevA.84.052721}, the $R$-matrix propagation method combined with
the SVD approach is used. We divide the hyperradius into (N - 1) intervals with the set of grid points
$R_{1}< R_{2} < \cdots < R_{N}$. In the interval $[R_{i}, R_{i+1}]$, the SVD method is used to solve Eq.\;(\ref{3}).
With this solution, we can determine the R-matrix, which is defined as
\begin{align}
\label{16}
\underline{\mathcal{R}}(R)=\underline{\textsl{F}}(R)[\widetilde{\underline{\textsl{F}}}(R)]^{-1}\,,
\end{align}
where matrices $\underline{\textsl{F}}$ and $\widetilde{\underline{\textsl{F}}}$ can be calculated from the solution of Eqs.~(\ref{3}) and (\ref{9}) by
\begin{align}
\label{17}
F_{\nu,\upsilon'}(R)=\int d\Omega \Phi_{\nu}(R;\Omega)^{*}\psi_{\upsilon'}(R;\Omega)\,,
\end{align}
\begin{align}
\label{18}
\widetilde{F}_{\nu,\upsilon'}(R)=\int d\Omega \Phi_{\nu}(R;\Omega)^{*}\frac{\partial}{\partial R}\psi_{\upsilon'}(R;\Omega)\,.
\end{align}
Over the interval $[R_{1}, R_{2}]$, when the $\underline{\mathcal{R}}$ matrix at $R_{1}$ is known, the $\underline{\mathcal{R}}$ matrix at another point $R = R_{2}$ can be calculated as follows:
\begin{align}
\label{33}
\underline{\mathcal{R}}(R_2)=\underline{\mathcal{R}}_{22}
-\underline{\mathcal{R}}_{21}\left[\underline{\mathcal{R}}_{11}
+\underline{\mathcal{R}}(R_1)\right]^{-1}\underline{\mathcal{R}}_{12}.
\end{align}
Using the recurrence relation (\ref{33}) in the R-matrix propagation method, we can obtain the $R$-matrix at the matching point $R_m$ where the wave function is matched to the wave function in the asymptotic region, and the three-body system is one dissociated atom plus a bound two-body system.

The asymptotic wave function of the atom + dimer scattering process in $\tau$-set Jacobi coordinates can be written as
\begin{equation}
\label{asywf}
\psi_A^{(\lambda)}(\rho_1,\rho_2)=\sum_{i=1}^{N} \frac{\varphi_{i}\left(\rho_{1}^{\tau}\right) \mathcal{Y}_{l_{1}^{\tau} l_{2}^{\tau} J M}\left(\hat{\Omega}_{1}^{\tau}, \hat{\Omega}_{2}^{\tau}\right)\left[f_{\lambda}\left(k\rho_{2}^{\tau}\right) \delta_{i \lambda}-g_{i}\left(k\rho_{2}^{\tau}\right) K_{i \lambda}\right]}{\rho_{1}^{\tau} k\rho_{2}^{\tau}},
\end{equation}
where $\hat{\Omega}_{1}^{\tau}=\left(\theta_1^\tau,\phi_1^\tau \right)$ and $\hat{\Omega}_{2}^{\tau}=\left(\theta_2^\tau,\phi_2^\tau \right)$ are the orientation angles of the vectors $\mathbf{\rho_1}$ and $\mathbf{\rho_2}$, respectively, for the $\tau$ arrangement.
$\varphi_{i}(\rho_1^{\tau})$ are the wave functions of the dimer, and $f$ and $g$ are the energy-normalized regular and irregular spherical Bessel functions, respectively, and have the following form:
\begin{equation}
f_{\lambda}\left(k\rho_{2}^{\tau}\right)=\sqrt{\frac{2\mu_{\tau,\tau+1\tau+2}}{k\pi}}k\rho_{2}^{\tau}j_{\ell_{\tau,\tau+1\tau+2}}(k\rho_{2}^{\tau}),
\end{equation}
\begin{equation}
g_{i}\left(k\rho_{2}^{\tau}\right)=\sqrt{\frac{2\mu_{\tau,\tau+1\tau+2}}{k\pi}}k\rho_{2}^{\tau}n_{\ell_{\tau,\tau+1\tau+2}}(k\rho_{2}^{\tau}).
\end{equation}
The form of the angular part in the body frame is
\begin{equation}
\begin{aligned}
\mathcal{Y}_{l_{1}^{\tau} l_{2}^{\tau} J M}\left(\hat{\Omega}_{\mathrm{1}}^{\tau}, \hat{\Omega}_{2}^{\tau}\right)
&=\sqrt{\frac{16 \pi^{2}}{2 J+1}} \sum_{I}^{J} \frac{\tilde{D}_{I M}^{J\Pi}\left(\alpha, \beta, \gamma\right)}{1+(\sqrt{2}-1) \delta_{I0}} \mathcal{Y}_{l_{1}^{\tau} l_{2}^{\tau} JI}\left(\hat{\Omega}_{1}^{\tau(\mathrm{body})}, \hat{\Omega}_{2}^{\tau(\mathrm{body})}\right)\\
&=\sqrt{\frac{16 \pi^{2}}{2 J+1}} \sum_{I}^{J} \frac{\tilde{D}_{I M}^{J\Pi}\left(\alpha, \beta, \gamma\right)}{1+(\sqrt{2}-1) \delta_{I 0}} \sum_{m}\left\langle l_{1}^{\tau}\, m\, l_{2}^{\tau}\, I-m \mid J\, I\right\rangle \\
&\;\;\;\;~ \times Y_{l_{1}^{\tau} m}\left(\theta_{1}^{\tau(\mathrm{body})}, \phi_{1}^{\tau(\mathrm{body})}\right) Y_{l_{2}^{\tau} I-m}\left(\theta_{2}^{\tau(\mathrm{body})}, \phi_{2}^{\tau(\mathrm{body})}\right),
\end{aligned}
\end{equation}
where the superscript (body) means these angles are measured in the body-fixed frame.

For simplicity, $S_{Il_{1}^{\tau} l_{2}^{\tau} J M}\left( \theta^\tau,\phi^\tau\right)$ is introduced:
\begin{equation}
\begin{aligned}
S_{Il_{1}^{\tau} l_{2}^{\tau} J M}\left( \theta^\tau,\phi^\tau\right)\equiv\sqrt{\frac{16 \pi^{2}}{2 J+1}} \frac{1}{1+(\sqrt{2}-1) \delta_{I 0}}   \sum_{m}\left\langle l_{1}^{\tau}\, m\,l_{2}^{\tau}\,I-m \mid J\,I\right\rangle \\
 \times Y_{l_{1}^{\tau} m}\left(\theta_{1}^{\tau(\mathrm{body})}, \phi_{1}^{\tau(\mathrm{body})}\right) Y_{l_{2}^{\tau} I-m}\left(\theta_{2}^{\tau(\mathrm{body})}, \phi_{2}^{\tau(\mathrm{body}) }\right).
\end{aligned}
\end{equation}

After transforming the asymptotic wave function into a body frame, the matching process between the inner region wave function $\Psi=\sum_{\mu I} F_{\mu I}\left(R_{m}\right) u_{\mu I}\left(R_{m} ; \theta, \phi\right) \tilde{D}_{I M}^{J}\left(\alpha, \beta, \gamma\right) $ calculated in Delves coordinates and the asymptotic function $\psi_A$ in Jacobi coordinates can be implemented:
\begin{equation}
\label{matching}
\frac{1}{R_{m}^{5 / 2} \sin \phi \cos \phi} \sum_{\sigma=1}^{N} H_{\sigma}^{\lambda} \Psi^{(\sigma)}\left(R_{m},  \Omega\right)=\left.\psi_{A}^{(\lambda)}\left(\rho_{1}, \rho_{2}\right)\right|_{R=R_{m}},
\end{equation}
where $H_{\sigma}^{\lambda}$ denotes the expansion coefficients; that is,
\begin{equation}
\begin{aligned}
\label{matching2}
&\frac{\sum_{\mu I} \sum_{\sigma=1}^{N} H_{\sigma}^{\lambda} F_{\mu I}^{(\sigma)}\left(R_{m}\right) u_{\mu I}\left(R_{m} ; \theta, \phi\right) \tilde{D}_{I M}^{J\Pi}\left(\alpha, \beta, \gamma\right) }{R_{m}^{5 / 2} \sin \phi \cos \phi}\\
&=\sum_{i=1}^{N} \varphi_{i}\left(\rho_{1}^{\tau}\right)\left[f_{\lambda}\left(k\rho_{2}^{\tau}\right) \delta_{i \lambda}-g_{i}\left(k\rho_{2}^{\tau}\right) K_{i \lambda}\right]
 \frac{\sum_{I} \tilde{D}_{I M}^{J\Pi}\left(\alpha, \beta, \gamma\right) S_{Il_{1}^{\tau} l_{2}^{\tau} J M }\left(\theta^{\tau}, \phi^{\tau}\right)}{kR_{m}^{2} \cos \phi^{\tau} \sin \phi^{\tau} (\mu^{2} / \mu_{1}^{\tau} \mu_{2}^{\tau})^{ 1 / 2}}.
\end{aligned}
\end{equation}
Using the orthogonality and normalization of $u_{\mu I}\left(R_{m} ; \theta, \phi\right) \tilde{D}_{I M}^{J\Pi}\left(\alpha, \beta, \gamma\right)$, we can obtain the following relation:
\begin{equation}
\label{matching3}
\begin{aligned}
\sum_{\sigma=1}^{N} H_{\sigma}^{\lambda} F_{\mu I}^{(\sigma)}\left(R_{m}\right) &=\left.R_{m}^{5 / 2} \int(\sin \phi \cos \phi) u_{\mu I}\left(R_{m} ; \theta, \phi\right) \tilde{D}_{I M}^{J\Pi} \psi_{A}^{(\lambda)}\left(\rho_{1}, \rho_{2}\right)\right|_{R=R_{m}} \mathrm{d} \phi\mathrm{d} \hat{\Omega} \\
& \equiv R_{m}^{1 / 2}\left(J_{\mu I}^{\lambda}-\sum_{i=1}^{N} N_{\mu I}^{i} K_{i \lambda}\right),
\end{aligned}
\end{equation}
where
\begin{equation}
\begin{aligned}
\label{J}
J_{\mu I}^{\lambda}=\int &\frac{\sin \phi \cos \phi}{k\sin \phi^{\tau} \cos \phi^{\tau} (\mu^{2} / \mu_{1}^{\tau} \mu_{2}^\tau)^{1 / 2}} u_{\mu I}\left(R_{m} ; \theta, \phi\right) \\&\times\varphi_{\lambda}\left(\rho_{1}^{\tau}\right) f_{\lambda}\left(k\rho_{2}^{\tau}\right) S_{I l_{1}^{\tau} l_{2}^{\tau} J M }\left(\theta^{\tau}, \phi^{\tau}\right) \sin \theta \mathrm{d} \theta \mathrm{d} \phi,
\end{aligned}
\end{equation}
\begin{equation}
\begin{aligned}
\label{N}
N_{\mu I}^{i}=\int &\frac{\sin \phi \cos \phi}{k\sin \phi^{\tau} \cos \phi^{\tau} (\mu^{2} / \mu_{1}^{\tau} \mu_{2}^\tau)^{1 / 2}} u_{\mu I}\left(R_{m} ; \theta, \phi\right) \\&\times \varphi_{i}\left(\rho_{1}^{\tau}\right) g_{i}\left(k\rho_{2}^{\tau}\right) S_{I l_{1}^{\tau} l_{2}^{\tau} J M }\left(\theta^{\tau}, \phi^{\tau}\right) \sin \theta \mathrm{d} \theta \mathrm{d} \phi.
\end{aligned}
\end{equation}
At the matching point $R_m$, the logarithmic derivative of the inner and outer region wave functions should be equal; therefore, the derivative of the asymptotic wave function is also needed:
\begin{equation}
\begin{aligned}
\label{Jp}
J_{\mu I}^{\lambda \prime}=&\int \frac{\sin \phi \cos \phi u_{\mu I }\left(R_{m} ; \theta, \phi\right)}{k\sin \phi^{\tau} \cos \phi^{\tau}}\left(\sqrt{\frac{\mu_{2}^{\tau}}{\mu}} \cos \phi^{\tau} \varphi_{\lambda}^{\prime}\left(\rho_{1}^{\tau}\right) f_{\lambda}\left(k\rho_{2}^{\tau}\right)\right. \\
&\left.+\sqrt{\frac{\mu_{1}^{\tau}}{\mu}} \sin \phi^{\tau} \varphi_{\lambda}\left(\rho_{1}^{\tau}\right) f_{\lambda}^{\prime}\left(k\rho_{2}^{\tau}\right)\right)_{R=R_{m}} S_{Il_{1}^{\tau} l_{2}^{\tau} J M }\left(\theta^{\tau}, \phi^{\tau}\right) \sin \theta \mathrm{d} \theta \mathrm{d} \phi+\frac{J_{\mu I}^{\lambda}}{2 R_{m}}
\end{aligned}
\end{equation}
and
\begin{equation}
\begin{aligned}
\label{Np}
N_{\mu I}^{i \prime}=\int& \frac{\sin \phi \cos \phi u_{\mu I}\left(R_{m} ; \theta, \phi\right)}{k\sin \phi^{\tau} \cos \phi^{\tau}}\left(\sqrt{\frac{\mu_{2}^{\tau}}{\mu}} \cos \phi^{\tau} \varphi_{i}^{\prime}\left(\rho_{1}^{\tau}\right) g_i\left(k\rho_{2}^{\tau}\right)\right. \\
&\left.+\sqrt{\frac{\mu_{1}^{\tau}}{\mu}} \sin \phi^{\tau} \varphi_{i}\left(\rho_{1}^{\tau}\right) g_i^{\prime}\left(k\rho_{2}^{\tau}\right)\right)_{R=R_{m}} S_{Il_{1}^{\tau} l_{2}^{\tau} J M }\left(\theta^{\tau}, \phi^{\tau}\right) \sin \theta \mathrm{d} \theta \mathrm{d} \phi+\frac{N_{\mu I}^{i}}{2 R_{m}}.
\end{aligned}
\end{equation}

The matrix form of Eq.\;(\ref{matching3}) is
\begin{equation}
\label{F}
\textbf{F} \textbf{H}=R_{m}^{1 / 2}[\textbf{J}-\textbf{N} \textbf{K}].
\end{equation}
The derivative with respect to $R$ at the matching point $R_m$ is
\begin{equation}
\label{Fp}
\textbf{F}^{\prime} \textbf{H}=R_{m}^{1 / 2}[\textbf{J}{^\prime}-\textbf{N}{^\prime} \textbf{K}].
\end{equation}
According the definition of the R matrix,
$\textbf{R}=\textbf{F}\textbf{F}^{\prime -1}$, with (\ref{F}) and (\ref{Fp}), we can obtain the reaction matrix,
\begin{equation}
\textbf{K}=\frac{\textbf{J}-\textbf{RJ}^{\prime}}{\textbf{N}-\textbf{RN}^{\prime}},
\end{equation}
and the scattering matrix,
\begin{align}
\label{sk}
\textbf{S}=\frac{\mathbf{1}+i\textbf{K}}{\mathbf{1}-i\textbf{K}}\,,
\end{align}
where $\mathbf{1}$ is the $3\times 3$ unit matrix.
The relation between the atom + dimer scattering phase shift $\delta_0$ and the diagonal element scattering matrix $\textbf{S}$ is
\begin{equation}
S_{0 \leftarrow 0}^{0+}=\exp \left(2 i \delta_{0}\right).
\end{equation}
With the phase shift $\delta_0$, we can obtain the atom + dimer scattering length $a_{ad}$ through
\begin{equation}
a_{ad}=-\lim _{k_{\tau, \tau+1\tau+2} \rightarrow 0} \frac{\tan \delta_{0}}{k_{\tau, \tau+1\tau+2}},
\end{equation}
and the total cross section $\sigma_2$ is
\begin{equation}
\sigma_{2}=\sum_{J, \Pi} \sigma_{2}^{J \Pi}=\sum_{J, \Pi} \frac{(2 J+1) \pi}{k_{\tau, \tau+1\tau+2}^{2}}\left|S_{0 \leftarrow 0}^{J \Pi}-1\right|^{2}.
\end{equation}

\section{Results and discussion}
\subsection{Pair potentials}
\begin{figure}
 \centering
   \includegraphics[width=0.6\linewidth]{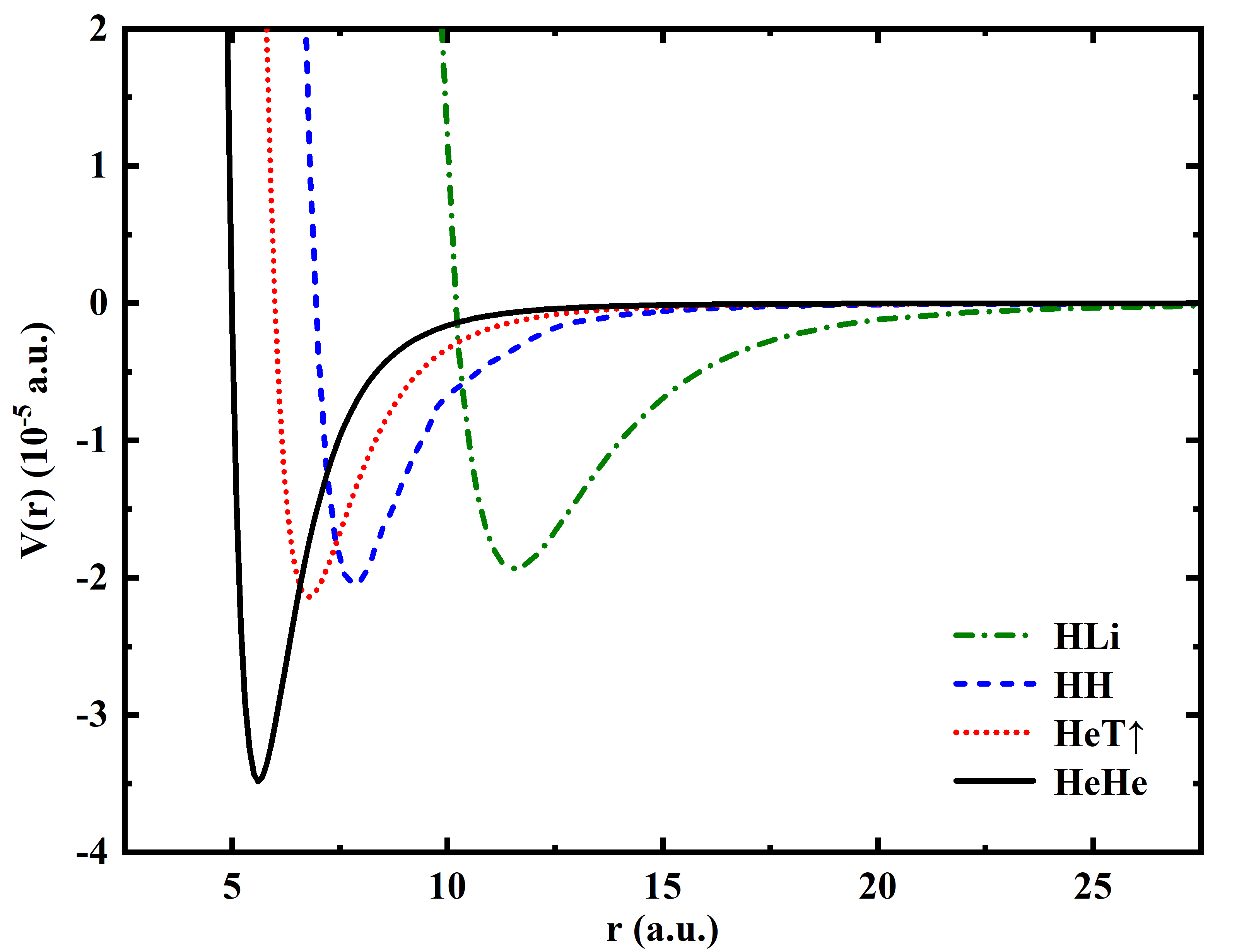}
\caption{Pairwise interaction potentials for He--He, He--T$\uparrow$, H--H, H--Li}
  \label{HH}
\end{figure}

For the helium dimer potential $v_{\textsl{HeHe}}(r)$, we use the CCSAPT potential of Jeziorska \textit{et al}.\;\cite{doi:10.1063/1.2770721}. The interaction between He and the spin-polarized tritium (T$\uparrow$) is identical to that between H and He. We choose the H--He potential developed by Cvetko \textit{et al}.\;\cite{doi:10.1063/1.466505}. The H and Li atoms are assumed to be spin-stretched. Their short-range potentials are determined from ab initio calculations\;\cite{doi:10.1063/1.1388044,doi:10.1063/1.1697142}, and their long-range behavior is determined by the usual dispersion potentials\;\cite{PhysRevA.71.032709,doi:10.1063/1.1388044}. All pairwise interaction potentials used in this work are shown in Fig.\;\ref{HH}. Their bound state energies $E_{\upsilon l}=E_{00}$ and scattering length $a$ calculated with potentials are summarized in Table \ref{t1}.

\begin{table}
\caption{Two-body bound state energies E$_{00}$ (in a.u.) and scattering length $a$ (in a.u.) for He--He, He--T$\uparrow$, H--H, and H--Li interactions.}
\renewcommand\arraystretch{0.55}
\begin{tabular*}{8.5cm}{@{\extracolsep{\fill}}lcc}
\hline\noalign{\smallskip}
   system                 &  E$_{00}$                &$a$ \\
\noalign{\smallskip}\hline\noalign{\smallskip}
$^{4}$He$_{2}$            &$-5.47114\times10^{-9}$   &165.5 \\
$^{4}$He$^{3}$He          &---------                 &-34 \\
$^{4}$He$^{3}$T$\uparrow$ &---------                 &-25 \\
H$\uparrow$-H$\uparrow$   &---------                 &1.6\\
Li-H$\uparrow$            &$-1.274\times10^{-7}$     &63.5\\
\hline\noalign{\smallskip}
\end{tabular*}
\label{t1}
\end{table}

\subsection{Matching in $A$-set: $^{3}$He$^{4}$He$_{2}$ and T$\uparrow^{4}$He$_{2}$ system}
In our definition of the hyperspherical coordinates, the two identical atoms are connected with Jacobi vector $\vec{\rho_1}$. Thus, the inner region wave functions are matched with the asymptotic solutions in $A$-set Jacobi coordinates for $^{3}$He$^{4}$He$_{2}$ and T$\uparrow^{4}$He$_{2}$ systems. Several papers have reported on the process of $^{3}$He atom scattering from the $^{4}$He$_{2}$ dimer, which is a good example to test our procedure. Kolganova and Sandhans \textit{et al}.\;\cite{2004Binding,2006Scattering,Kolganova2009Ultracold} calculated the scattering phase shifts and scattering length for $^{3}$He + $^{4}$He$_{2}$ using the two-dimensional partial-wave integral-differential Faddeev equations based on the SAPT2 and LM2M2 potentials. They estimated the scattering length to be between 35.9 a.u. and 37 a.u. based on these two kinds of potentials. Soon thereafter, Suno\;\cite{PhysRevA.78.062701} calculated the scattering length by solving the coupled-channel hyperradial equations using a combination of the finite element method\;\cite{Burke1999Theoretical}
and the R-matrix method\;\cite{RevModPhys.68.1015}. They used the improved He-He potential\;\cite{doi:10.1063/1.2770721} and predicted a $^{3}$He + $^{4}$He$_{2}$ scattering length value of 40 a.u..

Due to the similarities between $^{4}$He and T$\uparrow$ atoms, similar behaviors of mixed T$\uparrow$$^{4}$He$_{2}$ and $^{3}$He$^{4}$He$_{2}$ clusters are expected. For example, both systems have been found to possess one weakly bound state and exhibit a larger spatial extension with universal halo properties\;\cite{Hiroya2014A,PhysRevLett.113.253401,doi:10.1063/1.3530837}.
However, compared with the well-studied $^{3}$He + $^{4}$He$_{2}$ elastic scattering process, no scattering observables are available for the T$\uparrow$ atom scattering from the $^{4}$He$_{2}$ dimer.

\begin{figure}
 \centering
   \includegraphics[width=0.6\linewidth]{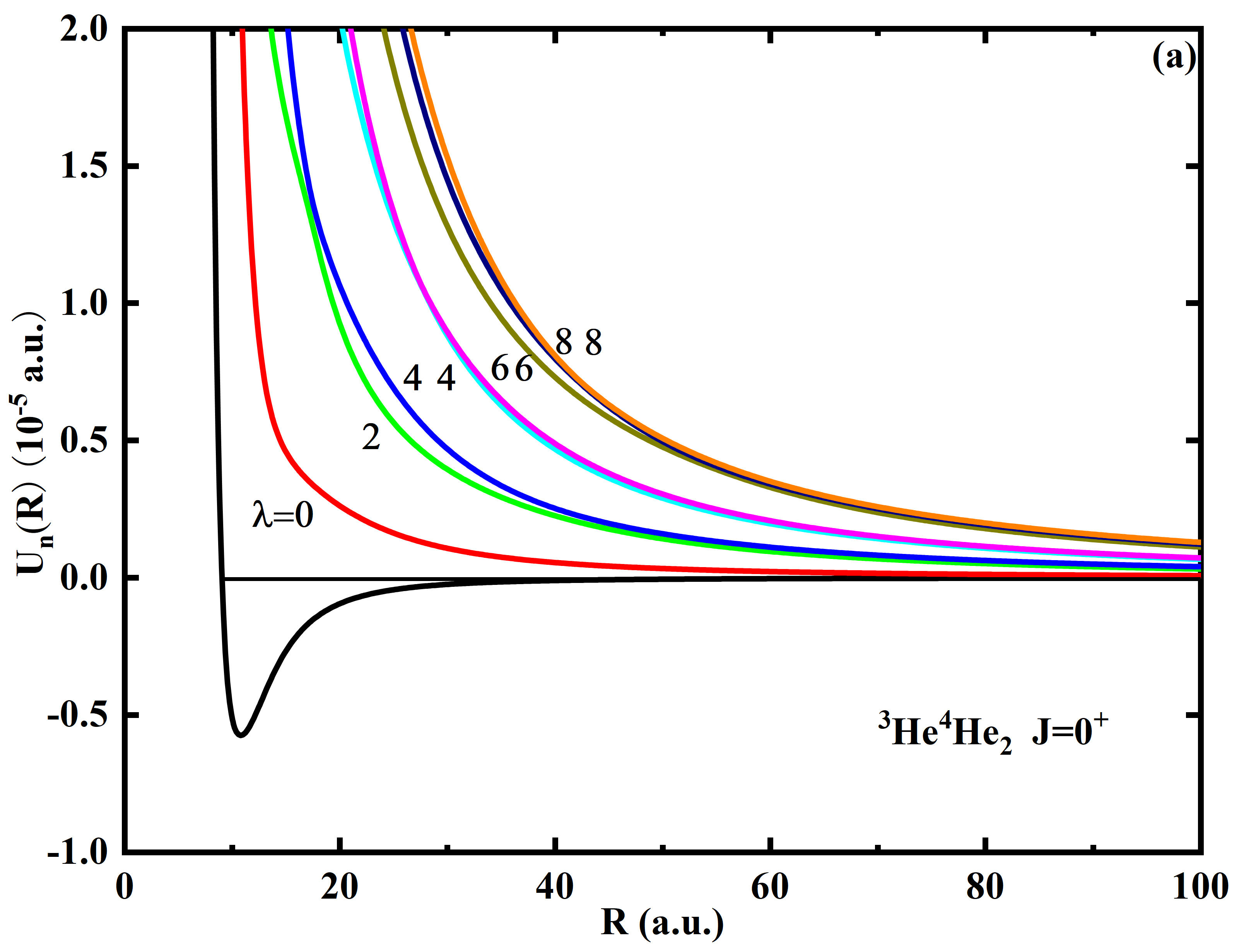}
    \includegraphics[width=0.6\linewidth]{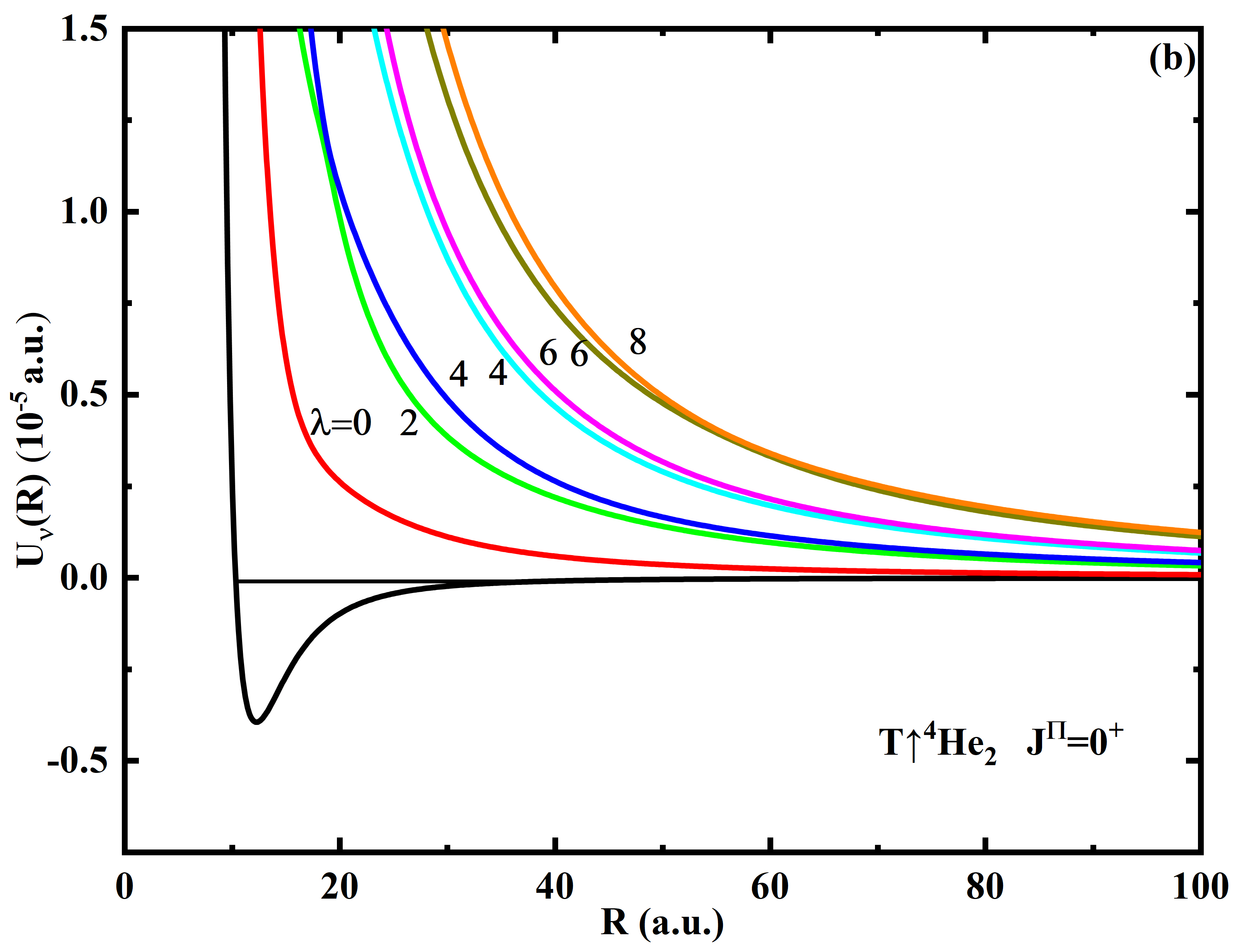}
\caption{Adiabatic hyperspherical potential curves $U_{\nu}(R)$ for (a) $^{3}$He$^{4}$He$_{2}$, $J^{\Pi}=0^{+}$ and (b) T$\uparrow$$^{4}$He$_{2}$, $J^{\Pi}=0^{+}$. The values indicate the asymptotic behavior of the potential curves, as given in Eq.\;(\ref{asywf}).}
  \label{f23}
\end{figure}

The potential curves of T$\uparrow$$^{4}$He$_{2}$ and $^{3}$He$^{4}$He$_{2}$ are presented in Fig.\;\ref{f23}. The lowest potential curves correspond asymptotically to the atom--dimer channel for T$\uparrow$-$^{4}$He$_{2}$ and $^{3}$He-$^{4}$He$_{2}$, and the other potential curves represent three-body continuum states. From Fig.\;\ref{f23}, the potential well of T$\uparrow$$^{4}$He$_{2}$ is shallower than that of the $^{3}$He$^{4}$He$_{2}$ system.
For ultracold atom--dimer collisions, the convergence of scattering observables depends critically on the accuracy of the adiabatic potentials. Thus, accurate potential curves and channel functions are highly desirable. According to the behavior of the channel function for these weakly bound systems, different B-spline knot distributions are used at short- and long-range hyperradii. For small hyperradii $R$, uniform knots are distributed; for large hyperradii $R$, the knot distribution is designed so that it becomes dense around the two-body coalescence points where the channel function is localized. Table\;\ref{t2} shows the convergence of lowest hyperspherical potential curves as functions of basis sets for T$\uparrow$$^{4}$He$_{2}$ and $^{3}$He$^{4}$He$_{2}$ systems. The basis sets $N_{\theta}=168$ and $N_{\phi}=504$ are chosen as the final calculation, and the potential curves have at least six significant digits. The convergence of the scattering observables with respect to the number of adiabatic channels and sectors is also tested. We typically use 13 channels and 230 sectors distributed as $R_{i}\propto i^{3}$ from $R=2$ a.u. to $R=500$ a.u..

\begin{table*}
\caption{Convergence test of adiabatic potentials (in a.u.) of T$\uparrow$$^{4}$He$_{2}$ and $^{3}$He$^{4}$He$_{2}$ as functions of the B-spline basis size ($N_{\theta}$, $N_{\phi}$). [a] abbreviates $\times10^a$. }
\label{t2}
\begin{tabular}{llllll}
\hline\noalign{\smallskip}
\multicolumn{6}{c}{T$\uparrow$$^{4}$He$_{2}$}  \\	
\noalign{\smallskip}\hline\noalign{\smallskip}
B($\theta$,$\phi$)&   R=20           &    R=100        &           R=200          &   R=300          &    R=500       \\
\hline\noalign{\smallskip}
(106, 304)         & -1.061490[-6]    &  -1.212912[-8]  &     -6.770491[-9]        & -6.030653[-9]    &  -5.668053[-9] \\
(106, 504)         & -1.061490[-6]    &  -1.212912[-8]  &     -6.770492[-9]        & -6.030666[-9]    &  -5.668068[-9] \\
(186, 504)         & -1.061490[-6]    &  -1.212912[-8]  &     -6.770492[-9]        & -6.030666[-9]    &  -5.668068[-9] \\
\hline\noalign{\smallskip}
\multicolumn{6}{c}{$^{3}$He$^{4}$He$_{2}$}  \\
\noalign{\smallskip}\hline\noalign{\smallskip}
$B(\theta,\phi)$   &   R=20             &    R=100          &           R=200              &   R=300            &    R=500                  \\
\hline\noalign{\smallskip}
(106, 304)          &   -9.431873[-7]    &   -1.222582[-8]   &		-6.782697[-9]      &   -6.032252[-9]    &   -5.668068[-9] \\
(106,404)           &   -9.431873[-7]    &   -1.222582[-8]   &      -6.782720[-9]      &   -6.032406[-9]    &   -5.668099[-9]  \\
(106, 504)          &   -9.431873[-7]    &   -1.222582[-8]   &		-6.782721[-9]      &   -6.032418[-9]    &   -5.668105[-9]    \\
(186, 504)          &   -9.431873[-7]    &   -1.222582[-8]   &		-6.782722[-9]      &   -6.032419[-9]    &   -5.668106[-9]    \\
\hline\noalign{\smallskip}
\end{tabular}	
\end{table*}

\begin{figure}
 \centering
   \includegraphics[width=0.6\linewidth]{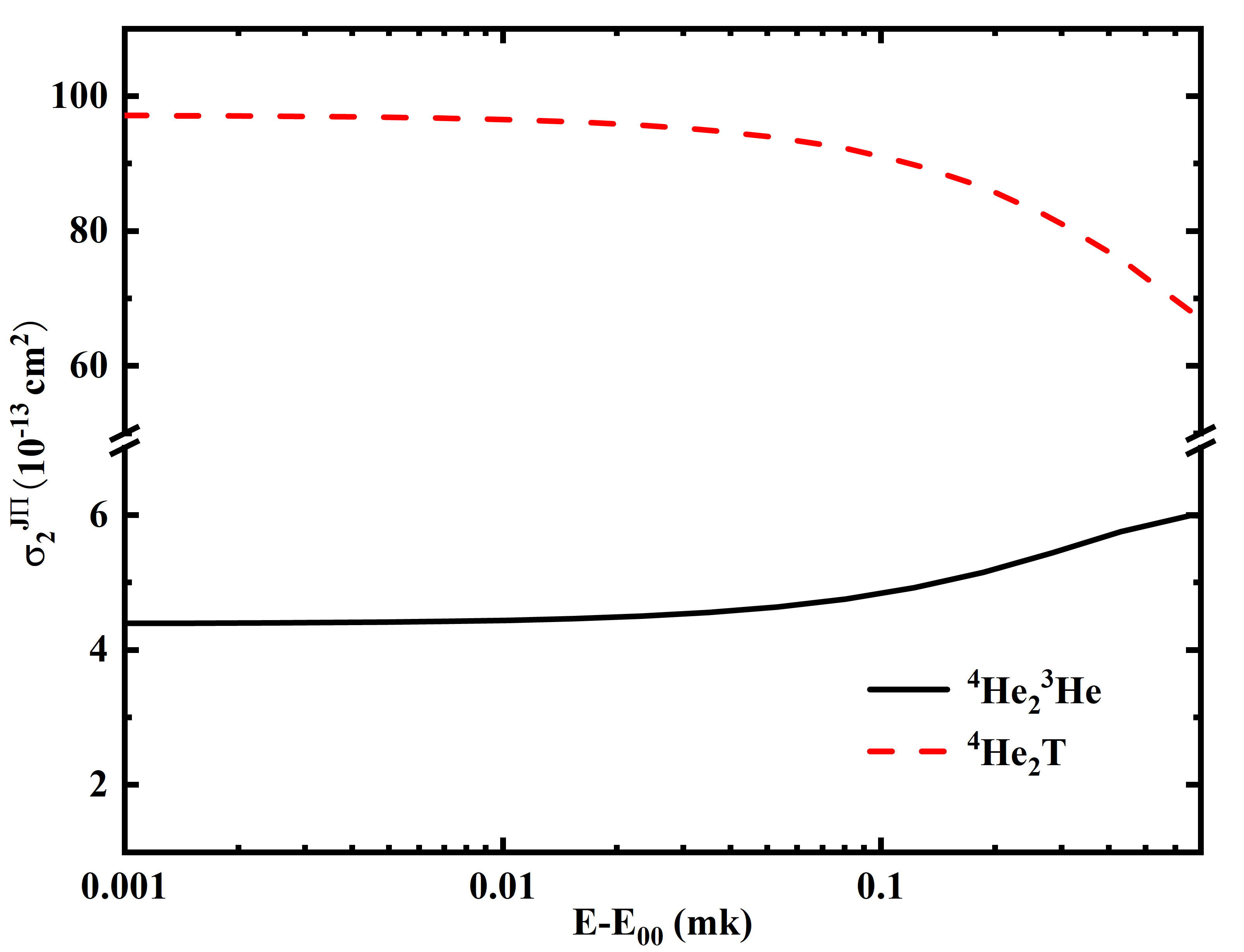}
\caption{$J=0$ cross sections for elastic $^{3}$He + $^{4}$He$_{2}$ and T$\uparrow$ + $^{4}$He$_{2}$ scattering as functions of the collision energy (E-E$_{00}$). }
  \label{f4}
\end{figure}

Figure \ref{f4} represents the $J=0$ cross sections for elastic $^{3}$He + $^{4}$He$_{2}$ and T$\uparrow$ + $^{4}$He$_{2}$ scattering as functions of the collision energy (E-E$_{00}$). In the ultracold limit, $\sigma^{0+}$ obeys the threshold behavior as $\sigma^{0+}\propto(E-E_{00})^{0}$.
Table\;\ref{t-3} shows the convergence of scattering lengths $a_{^{3}\textsl{He}+^{4}\textsl{He}_{2}}$ and $a_{\textsl{T}\uparrow+^{4}\textsl{He}_{2}}$ as a function of the matching distance. The scattering length converges at $R_{m}=500 \;a.u.$ for both systems. As shown in Refs.\;\cite{PhysRevA.78.062701,PhysRevA.83.032703}, the numerical solutions of these kinds of systems are usually matched to the asymptotic analytical solutions at $R_m = 5000\sim10000 $ a.u. in the hyperspherical coordinates boundary condition.

A comparison of our calculations with the results available in the literature is given in Table\;\ref{t-4}. For $^{3}$He + $^{4}$He$_{2}$ elastic scattering, Suno \textit{et al}.\;\cite{PhysRevA.78.062701} obtained $a_{^{3}\textsl{He} + ^{4}\textsl{He}_{2}} = 40\; a.u.\;$ using the potential from Ref.\;\cite{doi:10.1063/1.2770721}. With the same potential, the scattering length we calculated is $a_{^{3}\textsl{He} + ^{4}\textsl{He}_{2}} =34.6\; a.u.$. Sandhas \textit{et al}.\;\cite{Kolganova2009Ultracold} obtained $a_{^{3}\textsl{He} + ^{4}\textsl{He}_{2}}=37\; a.u.$ and $a_{^{3}\textsl{He} + ^{4}\textsl{He}_{2}} =35.9\;a.u.$ using LM2M2 and SAPT2 potentials, respectively.

For $^{3}$T$\uparrow$ + $^{4}$He$_{2}$ scattering, we obtain a scattering length value of $a_{\textsl{T}\uparrow+^4\textsl{He}_2} =166 \;a.u.$, which is larger than that of $^{3}$He + $^{4}$He$_{2}$ scattering. This result supports Suno's result that the $^{3}$T$\uparrow$ + $^{4}$He$_{2}$ bound state extends to larger distances than the $^{3}$He + $^{4}$He$_{2}$ bound state.

\begin{table}
\caption{Convergence test of the scattering lengths (in a.u.) of $a_{\textsl{T}\uparrow+^4\textsl{He}_2}$ and $a_{^3\textsl{He}+^4\textsl{He}_2}$ as functions of matching distances $R_{m}$(in a.u.).}
\label{t-3}
\renewcommand\arraystretch{0.55}
\begin{tabular*}{10cm}{@{\extracolsep{\fill}}ccc}
\hline\noalign{\smallskip}
   $R_{m}$  & $a_{^3\textsl{He}+^4\textsl{He}_2}$ &$a_{\textsl{T}\uparrow+^4\textsl{He}_2}$          \\
\noalign{\smallskip}\hline\noalign{\smallskip}
    200         & 40.9                     &                     \\
    300        & 35.2                       & 168.3               \\
    400       & 34.1                        & 166.6                \\
    500     &  34.6                         &166.8                 \\
\hline\noalign{\smallskip}
\end{tabular*}
\end{table}

\begin{table}
\caption{Atom--molecule zero-energy scattering lengths (in a.u.) of $a_{\textsl{T}\uparrow+^4\textsl{He}_2}$ and $a_{^3\textsl{He}+^4\textsl{He}_2}$ . Other calculations are also shown in the table for comparison.}
\label{t-4}
\begin{tabular*}{10cm}{@{\extracolsep{\fill}}lccc}
\hline\noalign{\smallskip}
          &He--He potential &  $a_{^3\textsl{He}+^4\textsl{He}_2}$ & $a_{\textsl{T}\uparrow+^4\textsl{He}_2}$\\
\noalign{\smallskip}\hline\noalign{\smallskip}
present &       CCSAPT\;\cite{doi:10.1063/1.2770721}           &          34                  &       166           \\
Ref.\;\cite{PhysRevA.78.062701}&   CCSAPT\;\cite{doi:10.1063/1.2770721}  & 40                  &                  \\
Ref.\;\cite{Kolganova2009Ultracold}                           & LM2M2\;\cite{doi:10.1063/1.460139}   & 37               &                     \\
Ref.\;\cite{Kolganova2009Ultracold}                           &  SAPT2\;\cite{doi:10.1063/1.474444}  &  35.9               &                       \\
\hline\noalign{\smallskip}
\end{tabular*}
\end{table}
\subsection{Matching in the $B$-set: LiHH system}

In the Delves hyperspherical coordinates defined in $A$-set Jacobi coordinates, where two identical atoms are connected with $\vec{\rho_1}$,
the asymptotic wave functions of the scattering process A + AC $\rightarrow$ A + AC (the A and C atoms are bounded) involve the transformation between the $A$-set and $B$-set. The spin-stretched case of H atom scattering from LiH is such an example where the lowest
adiabatic potentials asymptotically depend on the binding energies of the H--Li two-body bound states. Thus, the asymptotic wave function of the dissociated system is better represented in the $B$-set as follows:
\begin{equation}
\psi_A^{(\lambda)}(\rho_1^{B},\rho_2^{B})=\sum_{i=1}^{N} \frac{\varphi_{i}\left(\rho_{1}^{B}\right) \mathcal{Y}_{l_{1}^{B} l_{2}^{B} J M}\left(\hat{\Omega}_{1}^{B}, \hat{\Omega}_{2}^{B}\right)\left[f_{\lambda}\left(k\rho_{2}^{B}\right) \delta_{i \lambda}-g_{i}\left(k\rho_{2}^{B}\right) K_{i \lambda}\right]}{\rho_{1}^{B} k\rho_{2}^{B}}.
\end{equation}
The transformations between the $A$-set and $B$-set Jacobi coordinates can be implemented by the kinematic rotations given in Eq.\;(\ref{tfqt}).
The components of $\mathbf{\xi_2}$ and $\mathbf{\xi_1}$ in the $A$-set body frame can be written as
\begin{equation}
\left(\begin{array}{l}
\mathbf{\xi_2}_{A } \\
\mathbf{\xi_1}_{A }
\end{array}\right)=\left(\begin{array}{l}
\xi_{2A x}=\xi_{2A }\sin\theta \\
\xi_{2A  y}=0 \\
\xi_{2A  z}=\xi_{2A }\cos\theta \\
\xi_{1A  x}=0 \\
\xi_{1A  y}=0 \\
\xi_{1A  z}=\xi_{1A }
\end{array}\right).
\end{equation}
With Eqs. (\ref{tfq}) and (\ref{tfqt}), we can obtain the components of $\mathbf{\xi_2}$ and $\mathbf{\xi_1}$ in the $B$-set body frame as follows:
\begin{equation}
\begin{aligned}
\left(\begin{array}{l}
\mathbf{\xi_2}_{B} \\
\mathbf{\xi_1}_{B}
\end{array}\right)
=\left(\begin{array}{cc}
\cos \left(\chi_{B, A}\right) \mathbf{1} & \sin \left(\chi_{B, A}\right) \mathbf{1} \\
-\sin \left(\chi_{B, A}\right) \mathbf{1} & \cos \left(\chi_{B, A}\right) \mathbf{1}
\end{array}\right)
\left(\begin{array}{l}
\xi_{2A x}=\xi_{2A}\sin\theta \\
\xi_{2A y}=0 \\
\xi_{2A z}=\xi_{2A}\cos\theta \\
\xi_{1A x}=0 \\
\xi_{1A y}=0 \\
\xi_{1A z}=\xi_{1A}
\end{array}\right)
\end{aligned}
\end{equation}
and
\begin{equation}
\begin{aligned}
\left(\begin{array}{l}
\xi_{2 B x}=\cos \left(\chi_{B, A}\right)\xi_{2A}\sin\theta \\
\xi_{2 B y}=0 \\
\xi_{2 B z}=\cos \left(\chi_{B, A}\right)\xi_{2A}\cos\theta+\sin \left(\chi_{B, A}\right)\xi_{1A} \\
\xi_{1 B x}=-\sin \left(\chi_{B, A}\right)\xi_{2A}\sin\theta \\
\xi_{1 B y}=0 \\
\xi_{1 B z}=-\sin \left(\chi_{B, A}\right)\xi_{2A}\cos\theta +\cos \left(\chi_{B, A}\right)\xi_{1A}
\end{array}\right).\\
\end{aligned}
\end{equation}
With these equations, the expression of $\rho_{1B}$, $\rho_{2B}$,
$J_{\mu I}^{\lambda}$, $N_{\mu I}^{i}$, $J_{\mu I}^{\lambda \prime}$ and $N_{\mu I}^{i \prime}$ in $B$-set Jacobi coordinates can be obtained.

\begin{figure}
\centering
\includegraphics[width=0.6\textwidth]{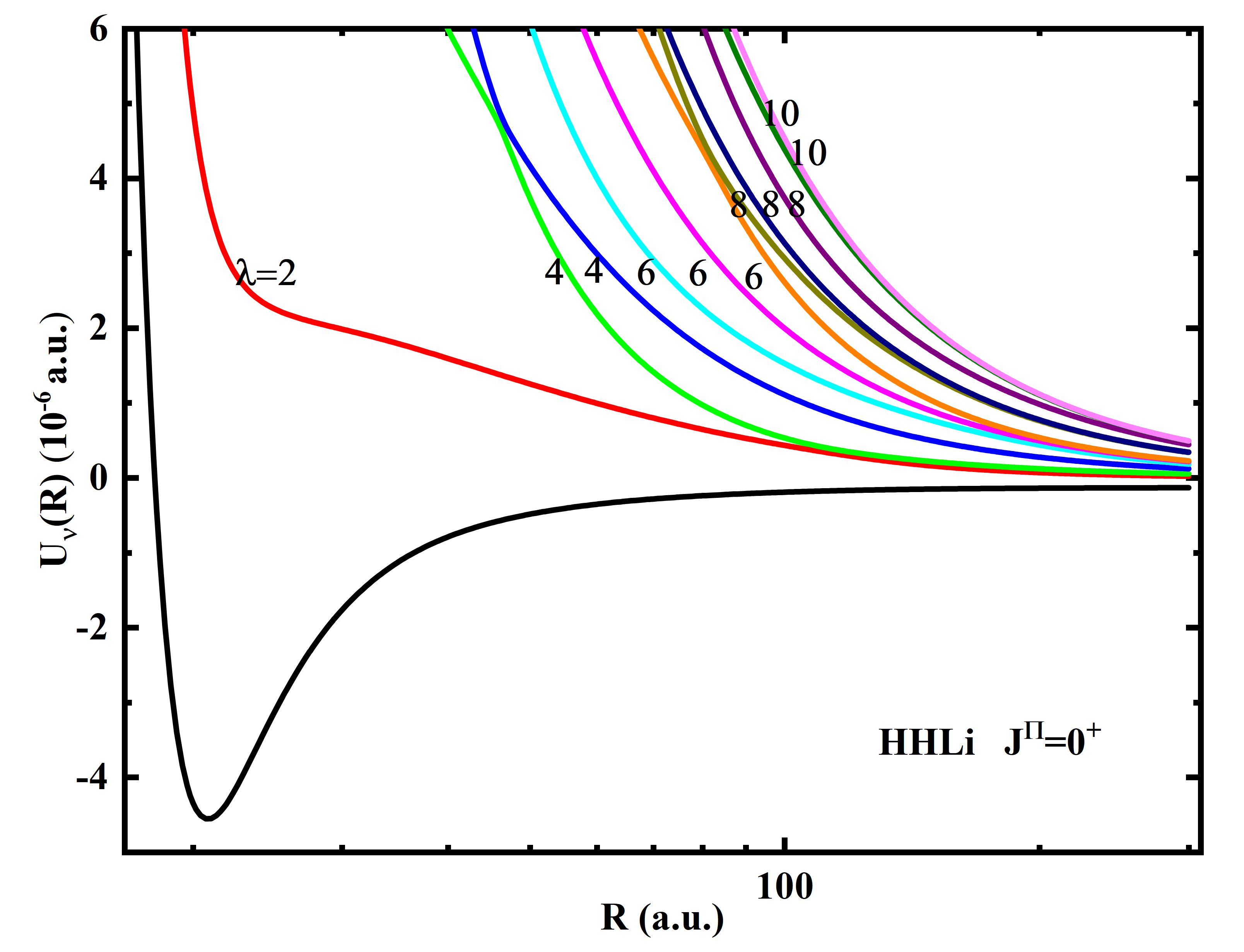}
\caption
{Adiabatic hyperpotential curves $U(R)$ for the H--H--Li system with $J^\Pi=0^+$. }
\label{f5-3}
\end{figure}

\begin{figure}
\centering
\includegraphics[width=0.6\textwidth]{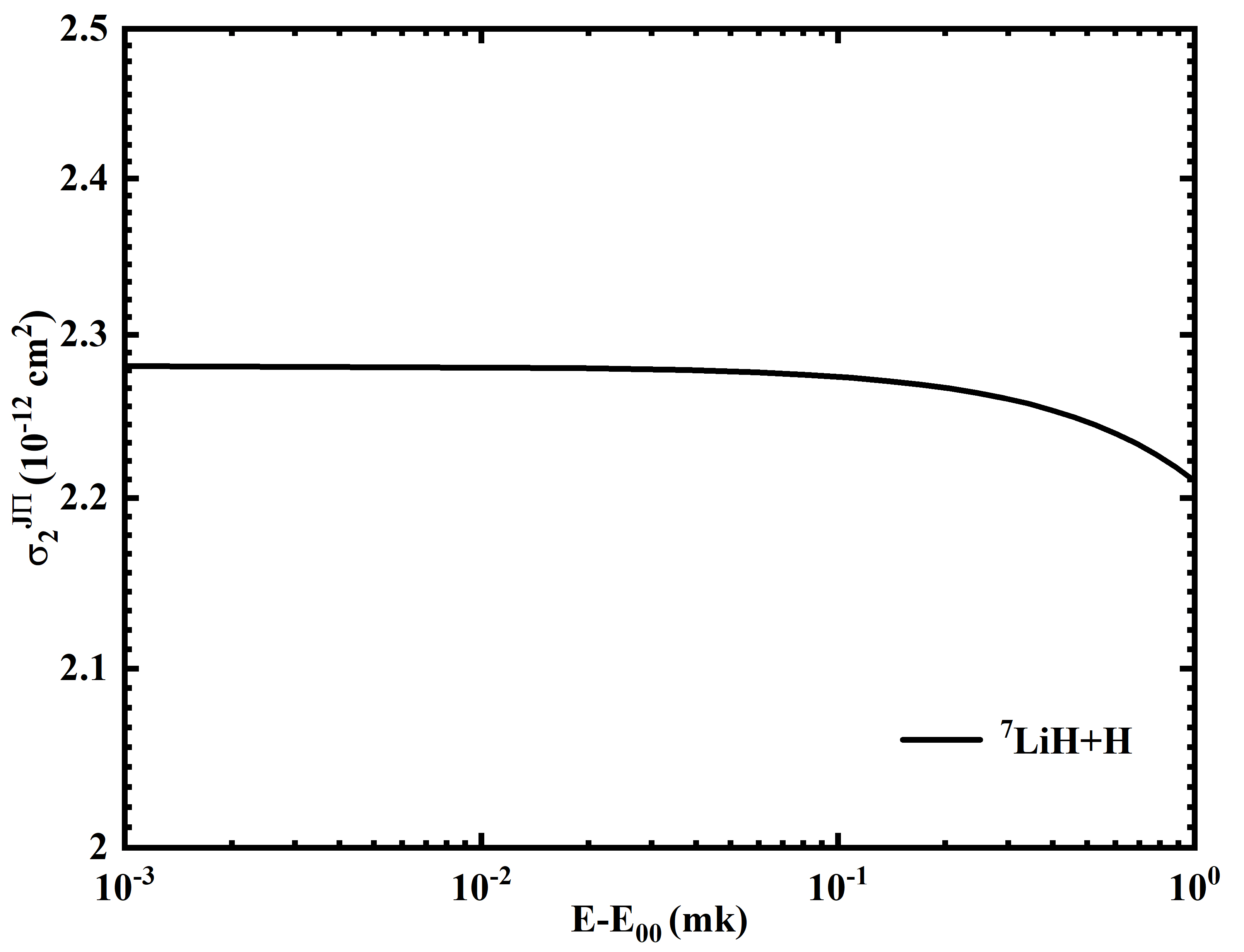}
\caption
{Cross section $\sigma_2$ for elastic H + LiH scattering as a function of the collision energy $(E-E_{00})$ for $J^\Pi=0^+$.  }
\label{f5-4}
\end{figure}

The adiabatic hyperspherical potentials $U_{\nu}(R)$ of the H--H--Li system are presented in Fig.\;\ref{f5-3}. The basis sets $N_{\theta}=168$ and $N_{\phi}=504$ are used, giving the potential curve at least six significant digits.
We plot the $J^{\Pi}=0^{+}$ partial wave cross sections for elastic collisions between H and LiH in Fig.\;\ref{f5-4}. For this system, 18 channels and 230 sectors are used to ensure that the scattering length has at least two digits.
Table\;\ref{t5-3} shows the convergence test of the atom--molecule scattering length $a_{\textsl{H+LiH}}$ as a function of the matching distance $R_{m}$. The scattering length is converged at the matching distance $R_m=500$ a.u. Note that Yujun \textit{et al}.\;\cite{PhysRevA.83.032703} calculated the elastic cross sections for H + LiH collisions in hyperspherical coordinate conditions. They matched the numerical solutions to the asymptotic analytical solutions at $R_m=5\times10^{3}$ a.u., and given the H + LiH scattering length of $80$ a.u. with the same $v_{HH}(r)$ and $v_{LiH}(r)$ potentials, the present result shows good agreement with their results.

\begin{table}
\caption{Convergence of the $J^\Pi=0^+$ scattering length between H and LiH $a_{\textsl{H+LiH}}$ (in a.u.) with respect to the matching distance $R_m$ (in a.u.).}
\label{t5-3}
\begin{tabular*}{8cm}{@{\extracolsep{\fill}}cccc}
\hline\noalign{\smallskip}
$R_m$ &$300$     &$400$    &$500$ \\
\noalign{\smallskip}\hline\noalign{\smallskip}
  $a_{\textsl{H+LiH}}$    &$79.7 $   &$80.9$    &$80.3 $ \\
\hline\noalign{\smallskip}
\end{tabular*}
\end{table}

\section{Conclusions}

In this work, we present an efficient method for solving the coupled-channel Schr\"odinger equation for atom--molecule elastic collisions. We use Delves hyperspherical coordinates, expand the wave function in a coupled-channel basis and propagate the coupled-channel equations with the R-matrix propagation technique. To avoid derivative coupling terms, we adopt the smooth variable discretization method, which discretizes the propagation variable before expanding in the basis. In the matching procedure, the asymptotic wave functions are expressed in the rotated Jacobi coordinates. Test calculations of elastic atom--molecule collisions are performed. For $^{3}$He(T$\uparrow$) atom scattering from $^{4}$He$_{2}$, the asymptotic wave functions are also expressed in the $A$-set Jacobi coordinates. No coordinate rotation is needed in this case. For the spin-stretched case of H atom scattering from LiH, the asymptotic wave functions must be expressed in the $B$-set Jacobi coordinates to describe the final scattering shape. Coordinate rotation between the $A$-set and $B$-set is needed for this type of scattering process. The convergence of the scattering length as a function of the propagation distance is studied. We find that the method is reliable and can improve the convergence as a function of matching distance. We compare our results with those of other calculations. The scattering length of H--HLi shows good agreement with that of hyperspherical coordinate boundary conditions with less computational expense. The scattering observables of T$\uparrow$-$^{4}$He$_{2}$ are scarce, whose scattering length and cross section values are given for the first time.

\begin{acknowledgements}
We thank C. H. Greene for helpful discussions. Hui-Li Han was
supported by the National Natural Science Foundation of China under Grant No. 11874391 and the National Key Research and Development
Program of China under Grant No. 2016YFA0301503.
Ting-Yun Shi was supported by the Strategic Priority Research Program of the Chinese Academy of Sciences under Grant No. XDB21030300.
\end{acknowledgements}

%
%
%


\end{document}